\documentclass[letterpaper,twocolumn,aps,floatfix]{revtex4}
\usepackage[latin1]{inputenc}
\usepackage{latexsym,amstext,amssymb,amsmath}
\usepackage[dvips]{graphicx}
\usepackage{color}
\begin{document}
%%%%%%%%%%%%%%%%

%%%%%%%%%%%%%%%%%%%%%%%%%%%%%%%%%%%%%%%%%%%%%%%%%%%%%%%%%%%%%%%%%
\title{Anomalous vibrational effects in non-magnetic and magnetic 
       Heusler alloys}   
%%%%%%%%%%%%%%%%%%%%%%%%%%%%%%%%%%%%%%%%%%%%%%%%%%%%%%%%%%%%%%%%%     

% \author{P. Entel}
% \email{entel@thp.uni-duisburg.de}
% \author{W. A. Adeagbo}
% \email{adeagbo@thp.uni-duisburg.de}
% \affiliation{Institute of Physics, University of Duisburg-Essen,
%             Duisburg Campus, 47048 Germany}
%
% \author{A. T. Zayak}
% \email{zayak@physics.rutgers.edu}
% \author{K. M. Rabe}
% \email{rabe@physics.rutgers.edu}
% \affiliation{Department of Physics and Astronomy, Rutgers University,
%  Piscataway, NJ 08854-8019, U.S.A.} 

 \author{A. T. Zayak$^{1,2}$}
 \email{zayak@physics.rutgers.edu}
 \author{P. Entel$^1$}
 \email{entel@thp.uni-duisburg.de}
 \author{K. M. Rabe$^2$}
 \email{rabe@physics.rutgers.edu}
 \author{W. A. Adeagbo$^1$}
 \email{adeagbo@thp.uni-duisburg.de}
 \author{M. Acet$^1$}
 \email{macet@agfarle.uni-duisburg.de}
 \affiliation{$^1$Institute of Physics, University of Duisburg-Essen,
             Duisburg Campus, 47048 Germany}
 \affiliation{$^2$Department of Physics and Astronomy, Rutgers University,
              Piscataway, NJ 08854-8019, U.S.A.} 
      
\date{\today}

%%%%%%%%%%%%%%%%
\begin{abstract}
%%%%%%%%%%%%%%%%
%
First-principles calculations are used in order to investigate 
phonon anomalies in non-magnetic and magnetic Heusler alloys. Phonon 
dispersions for several systems in their cubic L2$\mathrm{_1}$
structure were obtained along  
the [110] direction. We consider compounds which exhibit phonon 
instabilities and compare them with their stable counterparts. The
analysis of the electronic structure allows us to identify the characteristic
features leading to structural instabilities. The phonon dispersions
of the unstable compounds show that, while the acoustic modes tend to 
soften, the optical modes disperse in a way which is significantly 
different from that of the stable structures. The optical modes that
appear to disperse at anomalously low frequencies  
are Raman active, which is considered an indication of a 
stronger polarizability of the unstable systems. 
We show that phonon instability of the TA$_{2}$ mode in Heusler alloys
is driven by interaction(repulsion) with the low energy optical
vibrations. The optical modes show their unusual behavior 
due to covalent interactions which are additional bonding features
incommensurate with the dominating metallicity in Heusler compounds. 
%  
%%%%%%%%%%%%%%
\end{abstract}
%%%%%%%%%%%%%%

\maketitle

%%%%%%%%%%%%%%%%%%%%%%
\section{Introduction}
%%%%%%%%%%%%%%%%%%%%%%

In this paper, we present a detailed discussion of the electronic,
magnetic and vibrational properties of Heusler compounds. This subject is
of particular interest in relation to the magnetic shape-memory 
(MSM) effect recently discovered in these materials \cite{Takeuchi}.  
The Heusler compounds are known since 1903. However, only in the last 
decade have promising applications been presented for those systems 
which undergo a structural phase transformation and a magnetic phase 
transition near room temperature. The origin of the structural 
instability is related to the physics of martensitic transformations, 
of which the driving forces are not completely understood on a 
microscopic level \cite{Wasser}. Therefore, a detailed knowledge of the 
difference between structurally stable and unstable Heusler compounds 
on a microscopic scale would lead to a significant contribution to the 
theory of martensitic transformations in general and would be helpful 
when discussing further technical applications of MSM alloys.

%%%%%%%%%%%%%%%%%%%%%%%%%%%%%%%%%%%%%%%%%%%%%%%%%%%%%%%%%%%%%%%%%%%%%
\begin{figure} % Fig. 1a and b
  \begin{center}  
\includegraphics*[angle=0,width=8cm]{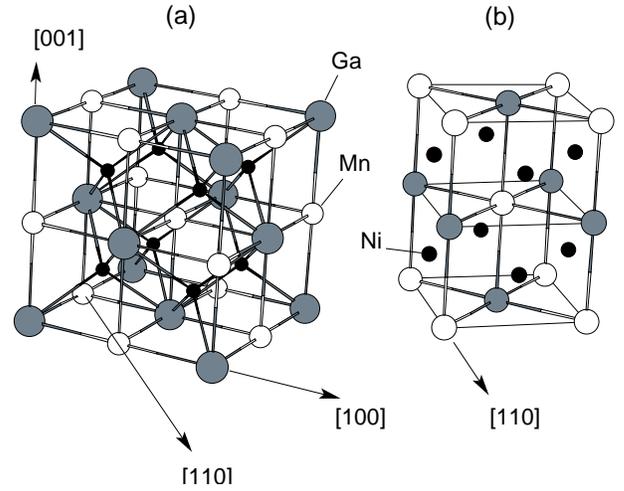}
  \end{center}
  \caption{(a) The L2$_1$ Heusler structure with 
  unit cell of space group Fm\,\=3\,m\,(O$_5^h$) of Ni$_2$MnGa
  \cite{Webster-84} and (b) conventional tetragonal cell used
  in the electronic structure calculations. All compounds considered
  in this work have the same structure and formula unit X$_2$YZ.}
  \label{struc}
\end{figure}
%%%%%%%%%%%%%%%%%%%%%%%%%%%%%%%%%%%%%%%%%%%%%%%%%%%%%%%%%%%%%%%%%%%%%

With respect to technological applications, Heusler alloys are
important for two reasons. Those which undergo a martensitic
transformation can be used to develop mechanical devices based on the
specific elastic properties of the martensitic structure, whereas the
ferromagnetically ordered martensites can be used for MSM technology
\cite{Ullakko-APL96}. The MSM physics is based on the fact that the
ferromagnetic martensitic structure can be deformed by applying
an external magnetic field, since in MSM systems, martensitic domains
are at the same time magnetic domains. It takes less energy to
redistribute the martensitic domains in case of strong uniaxial
magnetic anisotropy compared to the energy required to rotate the
local axis of magnetization. Corresponding domains that have
a favorable orientation of magnetization grow at the expense of domains
for which the magnetization directions are not parallel to the
external magnetic field. This has been experimentally observed in 
Ni$_2$MnGa \cite{Webster-84}, which today serves as a reference system 
for all investigations related to MSM technology \cite{Murray-00}. 
It has been shown that in moderate magnetic fields of the order of 1 T, the
structural deformations in Ni$_2$MnGa can reach 10\% \cite{Sozinov}.
This is the key feature of MSM technology that allows the
design of new kinds of micro-mechanical sensors and actuators.
Therefore, a complete understanding of the interplay of martensitic
micro-mechanics and magnetism is of prime technological interest.

In this paper, we present results of investigations of
the structural stability and electronic properties of the L2$_1$
ordered non-magnetic (NM), ferromagnetic (FM) and ferri-magnetic
(FerriM) Heusler compounds (the L2$_1$ structure is shown in Fig.\
\ref{struc}). From this comparative study, we expect to
obtain those critical parameters which distinguish unstable and
stable Heusler compounds. The present analysis is a continuation of previous
investigations of structural instabilities in such compounds,
whereby the instability has been related to Fermi surface nesting
in FM compounds \cite{Claudia,Harmon,Veliko-99}. In addition to the
nesting picture, we present in this paper further unusual features
of Heusler alloys which are related to characteristics of phonon 
dispersions and atomic displacements, electron and phonon densities of
states, and the coexistence of the metallic and covalent bondings.

%%%%%%%%%%%%%%%%%%%%%%%%%%%%%%%%%%%%%%%%%%%%%%%%%%%
\section{Heusler alloys selected for this study}
%%%%%%%%%%%%%%%%%%%%%%%%%%%%%%%%%%%%%%%%%%%%%%%%%%%

A set of eight Heusler alloys have been selected for this investigation. 
They are connected by a close relationship to Ni$_2$MnGa which is a
reference system for all studies related to the MSM
properties and Heusler alloys in general, exhibiting most of the
properties of interest. Moreover, we are quite
familiar with this system from the previous theoretical investigations
\cite{Claudia,Zayak-03,esomat}. At compositions close to
stoichiometric, this compound has the cubic L2$_\mathrm{1}$ structure
(see Fig. \ref{struc}) which is stable at high
temperature. Under cooling this crystal undergoes a series of magnetic
and structural phase transitions \cite{Martynov-92,Dai-2004}. 
Of central interest is the soft phonon mode shown by the
L2$_\mathrm{1}$ structure of 
Ni$_2$MnGa at low temperature \cite{Zheludev-95a}. The character of
this soft mode is believed to be related to the modulated structures
3M, 10M and 14M, coupling to a uniform tetragonal distortion and the
specific nesting topology of the Fermi surface of Ni$_2$MnGa
\cite{Claudia,Harmon,Zayak-03}. Similar properties have been found
in other Heusler compounds \cite{Thomas}. By making controlled changes
to the chemical composition of Ni$_2$MnGa we plan to explore chemical
trends in the electronic structure and structural energetics.

It can be helpful to characterize different Heusler compounds by their average 
concentration of valence electrons, which is defined as an
electron-to-atom ratio \cite{Chernenko-99,Chern-2002,Dai-2004}.
The parameter $e/a$ is a measure of the
electronic filling and allows in a rough way to distinguish in the
phase diagram of the ternary alloys the range of stability for the
martensitically unstable alloys. In
case of Ni$_2$MnGa, Ni contributes ten, Mn seven, and Ga three valence
electrons yielding $e/a = 7.5$. We consider those compounds which have
the same number of valence electrons as Ni$_2$MnGa and those whose $e/a$ are
slightly different. Thus, by substituting one atom of Ga in Ni$_2$MnGa
by atoms of Al or In, we obtain compounds Ni$_2$MnAl and Ni$_2$MnIn
with $e/a = 7.5$. Then, we add one valence electron in the system by
substituting Ge to the place of Ga(Al,In), and obtain compound
Ni$_2$MnGe with $e/a = 7.75$. In Ni$_2$MnGa and  Ni$_2$MnGe, we
substitute Ni by Co in order to consider compounds Co$_2$MnGa ($e/a =
7.00$) and  Co$_2$MnGe ($e/a = 7.25$). 
A crossover from the Ni-based to
the Co-based compounds is especially important due to their very
different properties. The Co-based compounds are known as half-metals
\cite{Pickett,Half}. Eventually, we obtain more or less
systematic sampling of different compounds which, from preliminary
analysis, are known to be ferromagnetic. In order to discuss the role of 
magnetic order \cite{Harmon} along with ferromagnets we consider one
ferri-magnetic and one non-magnetic systems, Fe$_2$MnGa and
Ni$_2$TiGa, respectively.

%For many other existing Heusler compounds see Ref.~\onlinecite{LB32c}.

%%%%%%%%%%%%%%%%%%%%%%%%%%%%%%%%%%%%%%%%%%%%%%%%%%
\section{Method}
%%%%%%%%%%%%%%%%%%%%%%%%%%%%%%%%%%%%%%%%%%%%%%%%%%

%%%%%%%%%%%%%%%%%%%%%%%%%%%%%%%%%%%%%%%%%%%%%%%%
\subsection{First principles calculations}
%%%%%%%%%%%%%%%%%%%%%%%%%%%%%%%%%%%%%%%%%%%%%%%%

The Vienna {\it Ab-initio} Simulation Package (VASP)
\cite{Kresse-96,Kresse-99} was used to perform the first-principles
calculations. The projector-augmented wave formalism (PAW) 
implemented in this package \cite{PAW,Kresse-99} leads to very accurate
results comparable to other all-electron methods. The electronic
exchange and correlation are treated within density functional 
theory by using the
generalized gradient approximation \cite{GGA}. The expansion of the
electronic wave-functions in plane waves was done using the ``High
Precision'' option in VASP, which corresponds to
the kinetic energy cutoff as high as 337.3 eV or more, depending on
the system. For the
total-energy calculations in the conventional tetragonal
unit cell shown in Fig.\ref{struc}(b), we used a high-density mesh of
12$\times$12$\times$10 special $k$-points for Brillouin-zone
integrations. For the larger supercell used in the phonon
calculations, described below, we used a  
10$\times$2$\times$8 mesh.

%%%%%%%%%%%%%%%%%%%%%%%%%%%%%%%%%%%%%%%%%
\subsection{Phonon calculations}
%%%%%%%%%%%%%%%%%%%%%%%%%%%%%%%%%%%%%%%%%

Phonon dispersions along [110] were calculated from first principles
for each system in the L2$_\mathrm{1}$ structure at the theoretical
lattice constant given in Table \ref{alloys}.  We used the direct
force constant method \cite{Parlinski,Parlinski_M}.  
The supercell was a 1$\times$5$\times$1 periodic supercell based on
the  conventional tetragonal cell with lattice parameters 
$a_t = b_t = a/\sqrt{2}$, $c_t = c$, shown in Fig.\ref{struc}(b).
This is an orthorhombic supercell with the long axis along [110],
containing ten consecutive (110) atomic planes
along the [110] direction. We considered in turn each Cartesian
displacement of each crystallographically independent atom by 0.03 \AA\
and computed the induced forces acting on all other atoms within the
supercell using the Hellmann-Feynman theorem. From these results, we
constructed the force constant matrix within the harmonic
approximation. The phonon frequencies and corresponding eigenvectors
were obtained by diagonalizing the corresponding dynamical matrix. 
The 1$\times$5$\times$1 supercell geometry allows the direct
calculation for five q-vectors along the [110] direction. These five
vectors satisfy equation 
$$
\exp{(2\pi \imath  \mathbf{q_L} \cdot \mathbf{L})} = 1,
$$
\noindent
where $\mathbf{L}$ denotes indices of the lattice constants in the
supercell (in our case from 0 to 4, giving points $\zeta =$ 0.0, 0.25, 0.5,
0.75, 1.0, for the normalized wave vector [$\zeta, \zeta, 0$], which
spans our Brillouin zone from its center to the boundary).
The interpolation of the dispersion curves is performed by setting the
force constants with range beyond five atomic planes (half the length
of our supercell) to zero; in the present case the decay of force
constants with distance is found to be quite rapid and thus this is an
excellent approximation.

%%%%%%%%%%%%%%%%%%%%%%%%%%%%%%%%%%%%%%%%%%%%%%%%%%
\section{Results}
%%%%%%%%%%%%%%%%%%%%%%%%%%%%%%%%%%%%%%%%%%%%%%%%%%

%%%%%%%%%%%%%%%%%%%%%%%%%%%%%%%%%%%%%%%
\subsection{L2$\mathbf{_1}$ structure and magnetic ordering}
%%%%%%%%%%%%%%%%%%%%%%%%%%%%%%%%%%%%%%%

%%%%%%%%%%%%%%%%%%%%%%%%%%%%%%%%%%%%%%%%%%%%%%%%%%%%%%%%%%%%%%%%%%%%%%
\begin{table*}
  \begin{center}
 \caption{Computed lattice 
 parameters, magnetic moments per unit cell and types of magnetic
 order for the eight selected Heusler compounds in the L2$_1$
 structure. `Instability' of the cubic structure means  
 here that an unstable mode appears in the calculated phonon
 dispersion. The valence-electron-to-atom ratio $e/a$ for the NM
 compound is shown in brackets in order to avoid direct comparison 
 with the magnetic systems in terms of this parameter.}
   \vspace{5mm}
   \begin{tabular*}
   {0.595\textwidth}
   {@{\extracolsep{\fill}}l c c c c c }
    \hline
\hline
    System  &  a$_{_\mathrm{L2_1}}$ (\AA) & $\mu_\mathrm{_{total}}$
    ($\mu_{_\mathrm{B}}$) 
    &  L2$_1$  &  Magn. order & $e/a$ \\ \hline
   Ni$_2$MnGa & 5.8067  & 4.35 &  unstable  &  FM & 7.50 \\
   Ni$_2$MnAl  & 5.7000 & 4.20 &  unstable  &  FM & 7.50 \\ 
   Ni$_2$MnIn  & 6.0624 & 4.22 &  unstable  &  FM & 7.50 \\ 
   Ni$_2$MnGe  & 5.8039 & 4.10 &  unstable  &  FM & 7.75 \\ 
   Co$_2$MnGa  & 5.7100 & 5.09 &  stable  &  FM & 7.00\\
   Co$_2$MnGe  & 5.7285 & 4.99 &  stable  &  FM & 7.25 \\ 
   Ni$_2$TiGa  & 5.8895 & 0.00 &  unstable  &  NM & (6.75) \\
   Fe$_2$MnGa  & 5.6882 & 2.15  &  stable  &  FerriM & 6.50 \\ \hline
  \end{tabular*}
   \label{alloys}
  \end{center}
\end{table*}
%%%%%%%%%%%%%%%%%%%%%%%%%%%%%%%%%%%%%%%%%%%%%%%%%%%%%%%%%%%%%%%%%%%%%%%

For the stoichiometric Heusler compounds (X$_2$YZ) 
considered in this work, the calculated 
lattice constants, magnetic moments per formula unit and the nature 
of magnetic ordering in the L2$_1$ structure are given in Table
\ref{alloys}. All atoms occupy high symmetry positions within the
Fm\=3m (\#225) symmetry, where Wyckoff positions are 4a (0,0,0) for Y atoms; 4b
(1/2, 1/2, 1/2) for Z atoms; 8c (1/4, 1/4, 1/4) for X atoms.
The computed lattice constants largely agree with
the experimental values, where available, as summarized in Ref.~\onlinecite{LB19c,LB32c}. 
The magnetic moment of Fe$_2$MnGa is the result of 
local magnetic moments of Mn and Fe being aligned
anti-ferrimagnetically. Further discussions of the magnetic 
properties of the Heusler alloys can be found in
Ref.~\onlinecite{Kuebler83,SP-Galanakis}.

%%%%%%%%%%%%%%%%%%%%%%%%%%%%%%%%%%%%%%%
\subsection{Calculated phonon dispersions}
%%%%%%%%%%%%%%%%%%%%%%%%%%%%%%%%%%%%%%%

%%%%%%%%%%%%%%%%%%%%%%%%%%%%%%%%%%%%%%%%%%%%%%%%%%%%%%%%%%%%%%%%%%%
\begin{figure*} % Fig. 4
  \begin{center}  
%  \includegraphics*[angle=0,width=4.24cm]{Ni2MnGa_left.eps}
%\includegraphics*[angle=0,height=7.3cm]{figure4.eps}\hspace{-1.5mm}
%  \includegraphics*[angle=0,width=3.76cm]{Ni2MnAl_right.eps}
%\includegraphics*[angle=0,height=7.3cm]{figure5.eps}\hspace{-1.5mm}
%  \includegraphics*[angle=0,width=4.21cm]{Ni2MnGe_left.eps}
%\includegraphics*[angle=0,height=7.3cm]{figure6.eps}\hspace{-1.5mm}
%  \includegraphics*[angle=0,width=3.79cm]{Ni2MnIn_right.eps}
%\includegraphics*[angle=0,height=7.3cm]{figure7.eps}\hspace{-1.5mm}\\
%\vspace{1cm}
%  \includegraphics*[angle=0,width=4.24cm]{Co2MnGa_left.eps}
%\includegraphics*[angle=0,height=7.8cm]{figure8.eps}\hspace{-1.5mm}
%  \includegraphics*[angle=0,width=3.75cm]{Co2MnGe_right.eps}
%\includegraphics*[angle=0,height=7.8cm]{figure9.eps}\hspace{-1.5mm}
%  \includegraphics*[angle=0,width=4.24cm]{Ni2TiGa_left.eps}
%\includegraphics*[angle=0,height=7.8cm]{figure10.eps}\hspace{-1.5mm}
%  \includegraphics*[angle=0,width=3.75cm]{Fe2MnGa_right.eps}
%\includegraphics*[angle=0,height=7.8cm]{figure11.eps}\hspace{-1.5mm}
\includegraphics*[angle=0,width=\textwidth]{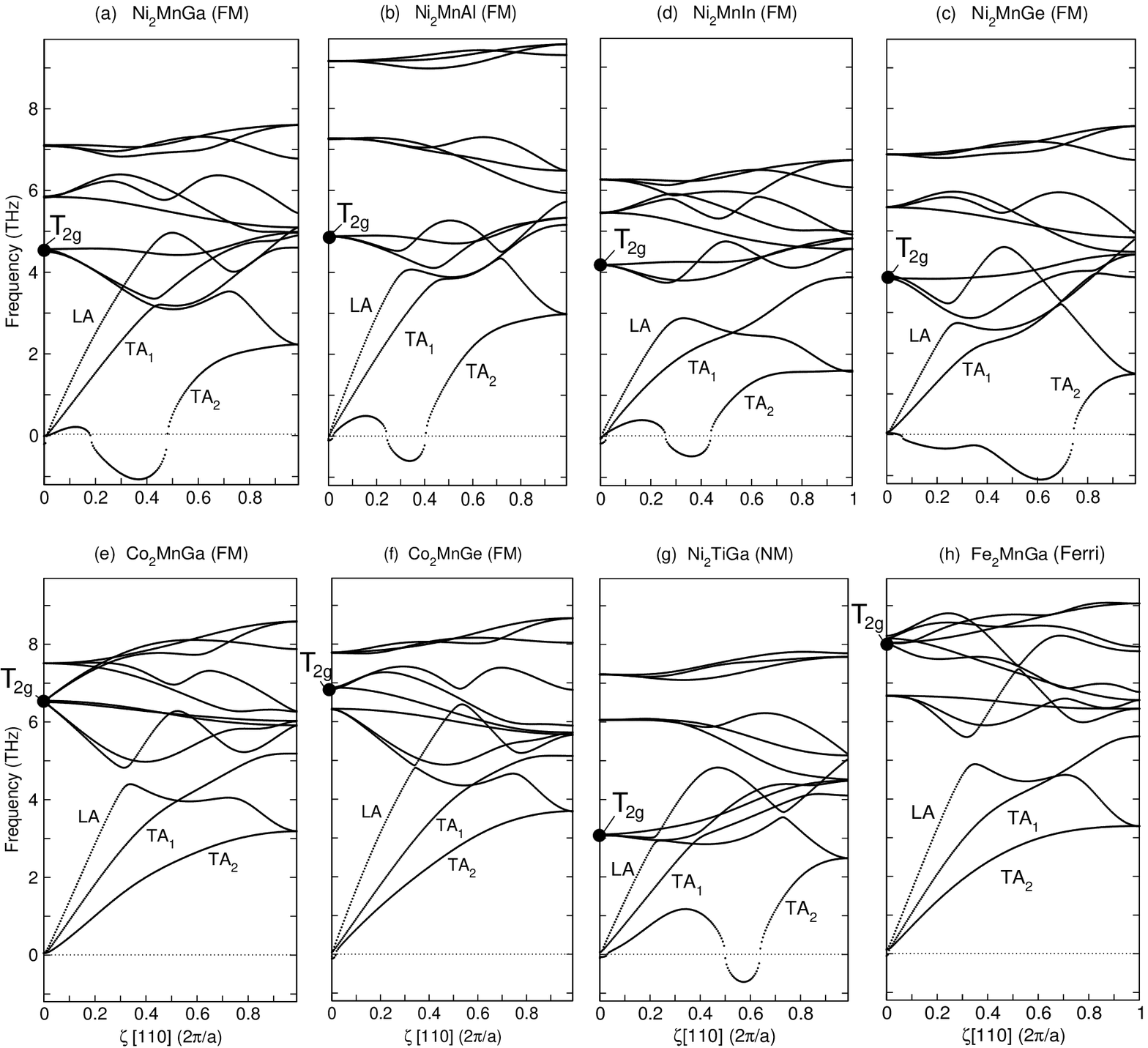}
  \end{center}
  \caption{Phonon dispersion curves of (a) FM Ni$_2$MnGa,  
  (b) FM Ni$_2$MnAl, (c) FM Ni$_2$MnGe, (d) FM Ni$_2$MnIn (e) FM Co$_2$MnGa, 
  (f) FM Co$_2$MnGe, (g) NM Ni$_2$TiGa and (h) FerriM Fe$_2$MnGa in the L2$_1$
  structure. Here, 
  the reduced wave vector coordinate $\zeta$ spans the
  $fcc$ Brillouin zone from $\Gamma$ to $X$. Imaginary frequencies of the unstable
  modes are shown in the real negative frequency range. 
  The frequency of
  the optical modes T$_{2g}$ at $\Gamma$ is marked by a black; note 
  it appears in lower values as compared to the
  stable systems. 
} 
  \label{g1}
\end{figure*}
%%%%%%%%%%%%%%%%%%%%%%%%%%%%%%%%%%%%%%%%%%%%%%%%%%%%%%%%%%%%%%%%%%%

The calculated phonon dispersions of the eight selected Heusler
compounds in their theoretical cubic L2$_\mathrm{1}$ structures  
are shown in Fig. \ref{g1}. Some of the phonon dispersions, shown in 
Fig. \ref{g1} (Ni$_2$Mn(Ga,Al)), have already been discussed in recent publications 
\cite{Zayak-PRB,Thomas}. Comparison of the phonon dispersions 
of Ni$_2$MnGa with existing experimental data and with phonon
dispersion calculations by other groups employing the linear response 
method show that the direct method used here yields fairly
accurate results comparable to the former method
\cite{Claudia,Zheludev-96}.  

The [110] dispersions of the L2$_1$ structure have one set of
nondegenerate acoustic branches (L, TA$_1$ and TA$_2$) and three optical
branches that can be easily recognized in Fig.~\ref{g1}. For the five
compounds Ni$_2$Mn(Ga,Al,In,Ge) and Ni$_2$TiGa, the TA$_2$ branch is
unstable for some range of $\zeta$. In 
addition, in Ni$_2$MnGe the TA$_2$ mode has a negative slope at $\Gamma$,
indicating a pure elastic instability. The instability of the L2$_1$
structure in NM Ni$_2$TiGa shows  
that magnetic order is not a necessary condition for the phonon 
softening to occur. 

%The upper acoustic branches and the lower optical branches clearly mix
%for $\zeta \approx 0.4$. Therefore, we also have to take  
%into account the effect of wave vector dependent anti-crossing, 
%which arises from the fact 
%that optical and acoustic phonon modes with the same symmetry are 
%not allowed to cross. An example of such a situation is shown in Fig.\ \ref{g2}.  
%The repulsion, which appears from the
%acoustic-optical interaction, lowers the frequencies of acoustic modes. 
%At the $\Gamma$ point, the acoustic and optical modes have well defined
%different symmetries. However, with increasing wave vector, the difference
%vanishes leading to a stronger repulsion of the modes.

\begin{figure} % Fig. 3
  \begin{center}  
\includegraphics*[angle=0,width=7.0cm]{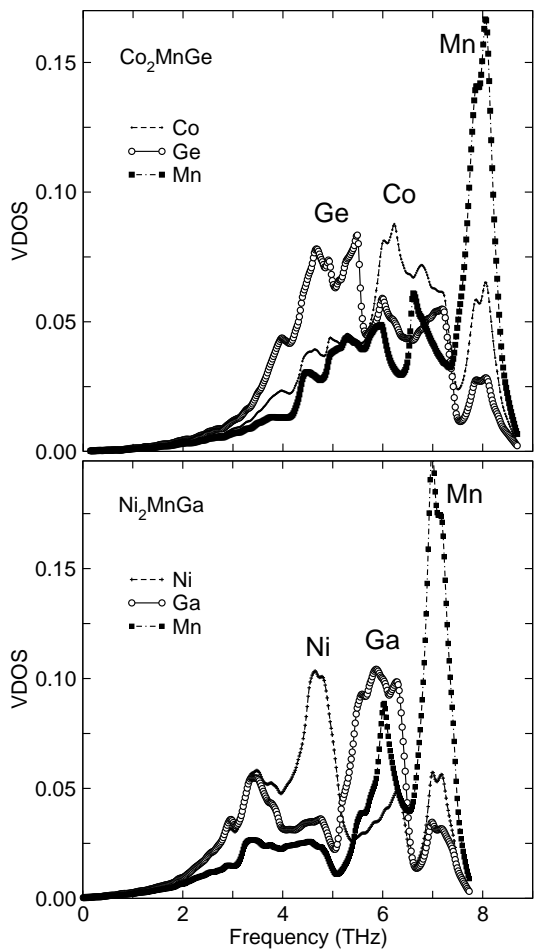}
  \end{center}
  \caption{Site projected vibrational density of states showing the
  distribution of amplitudes of the atoms over the frequency range. 
  This information shows which atom
  dominates in the vibrations of a given frequency range. In particular, 
  the anomalous behavior of optical vibrations of Ni lying 
  at frequencies below those of the the heavier Ga atom in unstable 
  Ni$_2$MnGa is clearly visible (see also Fig.\ \ref{g1}). 
  In the case of the stable system Co$_2$MnGe, the sequence 
  of the optical modes is regular, i.e. increasing as the mass of the
  atom decreases.}  
  \label{vdos}
\end{figure}

For some of the Heusler compounds the optical modes are split into
three well separated groups which are triply degenerate at $\Gamma$, but
become more mixed with increasing wave vector. In each of these three
groups 
one finds dominating vibrations of a particular kind of atom. Mostly,
this is due to the differences in the atomic masses and it can nicely
be established by computing
the partial phonon density of states for the [110] branch, shown for
two selected compounds in Fig. \ref{vdos}. 

Comparison of these two
plots shows that the position of the Ni-peak in case of unstable
compound Ni$_2$MnGa is different from the position of the Co-peak in
stable system Co$_2$MnGe.
This was unexpected, because atoms of Ni and Co, in both compounds,
occupy the same sites in the structure, Wyckoff positions 8c(1/4, 1/4,
1/4). Their masses are close to each other. Thus, the sequence of the
optical modes should be similar. However, while in case of stable
compound Co$_2$MnGe the sequence of the optical modes is normal, for
the case of Ni$_2$MnGa we observe an inversion of the optical modes.

%%%%%%%%%%%%%%%%%%%%%%%%%%%%%%%%%%%%%%%%%%%%%%%%%%%%%%%%%%%%%%%%%%%%%%

%%%%%%%%%%%%%%%%%%%%%%%%%%%%%%%%%%%%
\section{Discussions}
%%%%%%%%%%%%%%%%%%%%%%%%%%%%%%%%%%%%

%%%%%%%%%%%%%%%%%%%%%%%%%%%%%%%%%
\subsection{Inversion of the optical modes}
%%%%%%%%%%%%%%%%%%%%%%%%%%%%%%%%%

In the following we will discuss the impact of the inversion of the
optical modes shown in Fig.~\ref{vdos} in relation to the  
traditionally accepted point of view of the nature of  
incommensurate instabilities in Heusler alloys. As a matter of fact  
the structural instability of the L2$_1$ phase
is associated with  Fermi surface
nesting which also causes instability of the TA$_2$ mode 
in some systems leading to an incommensurate product phase, 
which has a modulated structure. The relation between soft mode 
behavior and specific nesting feature of the Fermi surface
has been recently emphasized and is not a debate any more
\cite{Claudia,Harmon,Veliko-99}.

%%%%%%%%%%%%%%%%%%%%%%%%%%%%%%%%%%%%%%%%%%%%%%%%%%%%%%%%%%%%%%%%%%%%%%
\begin{figure} % Fig. 3
  \begin{center}  
\includegraphics*[angle=0,width=6.0cm]{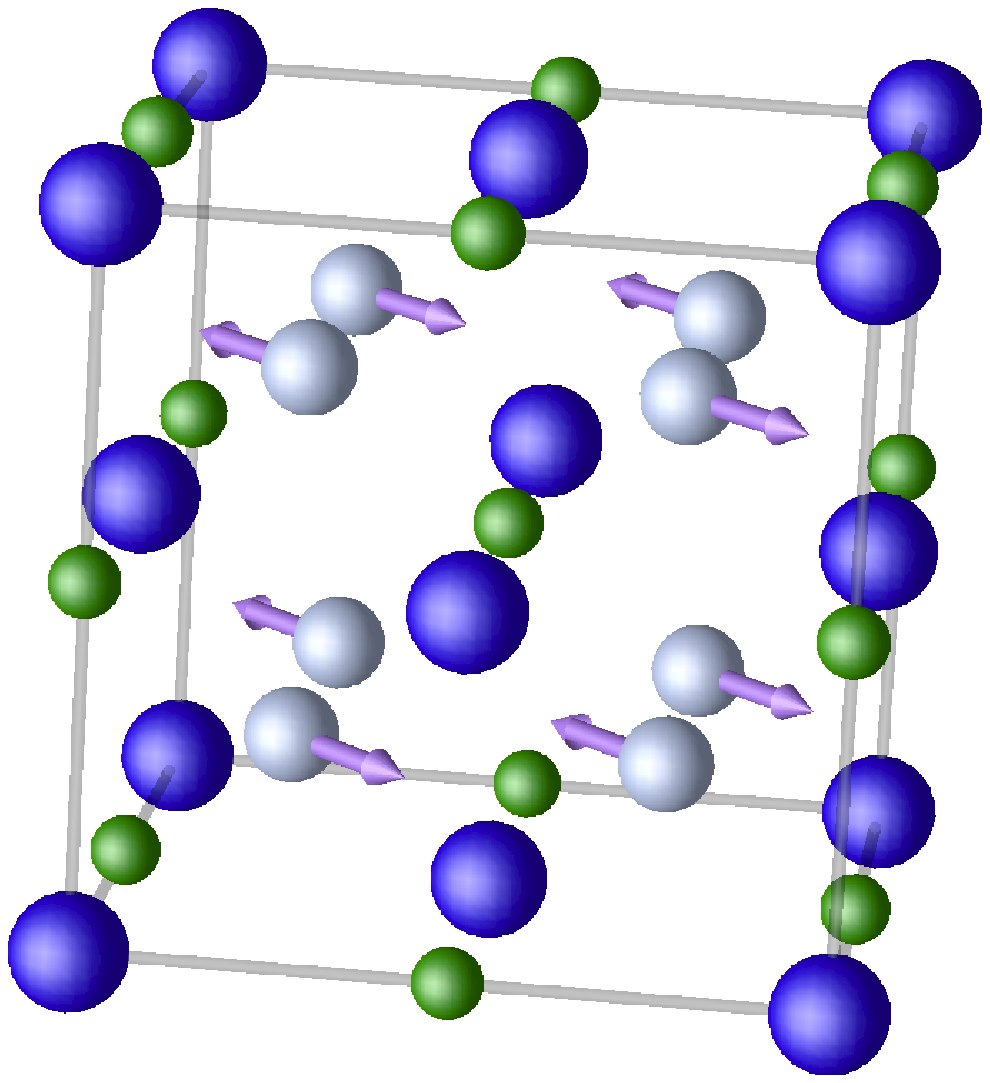}
  \end{center}
  \caption{
(Color online) The $\Gamma$-point T$_{2g}$ optical mode which vibrates
at anomalously 
low energy. Note that this is the only phonon mode with T$_{2g}$
symmetry in this structure and it is Raman active. 
Moreover, this mode is pure Ni-vibrations, of Co, i.e. the atoms
sitting in the 8c Wyckoff positions. Other atoms are not involved in
this mode. Other optical modes
involve vibrations of all atoms together and have symmetry T$_{1u}$ at
$\Gamma$, being IR-active.}   
\label{mode}
\end{figure}
%%%%%%%%%%%%%%%%%%%%%%%%%%%%%%%%%%%%%%%%%%%%%%%%%%%%%%%%%%%%%%%%%%%%%%

However, a systematic review of the phonon dispersions reveals a 
characteristic behavior of the optical T$_{2g}$ mode at $\Gamma$ (see
Fig.~\ref{mode}).
In all Ni-based systems we find that this mode appears at much lower
energies as compared to the Co-based stable compounds.  
The role of this anomaly might be as important as the instability of
the TA$_2$ mode, or in other words, these two features have to be
considered together in order to explain the structural instabilities
in Heusler alloys. 

We drew attention to the fact that phonon modes of the same
symmetry must repel each other. This is what must happen in case of 
the TA$_2$ and the T$_{2g}$ modes. At finite wave vectors only one optical mode
which has [1-10] polarization will repel with the the TA$_2$.
The main point is that there are two instabilities in the system. One of them
is that the (110) planes in the Heusler structure can slide in the [1-10]
direction, this is the TA$_2$ mode. But at the same time the (111)
planes of Ni slide against each other also along the same [1-10]
direction, and this is the T$_{2g}$ mode, shown in Fig.~\ref{mode}. 
These vibrations are destructive for each other and have to
repulse. Thus, the TA$_2$ mode is unstable because it is pushed down
by the corresponding from the symmetry optical T$_{2g}$ mode of Ni.
The Fermi surface nesting determines at which wave vector the TA$_2$
mode is most sensitive to the influence of the optical mode.

%%%%%%%%%%%%%%%%%%%%%%%%%%%%%%%%%
\subsection{Hybridization features}
%%%%%%%%%%%%%%%%%%%%%%%%%%%%%%%%%

\begin{figure} % Fig. 11
  \begin{center}  
\includegraphics*[angle=0,width=7.0cm]{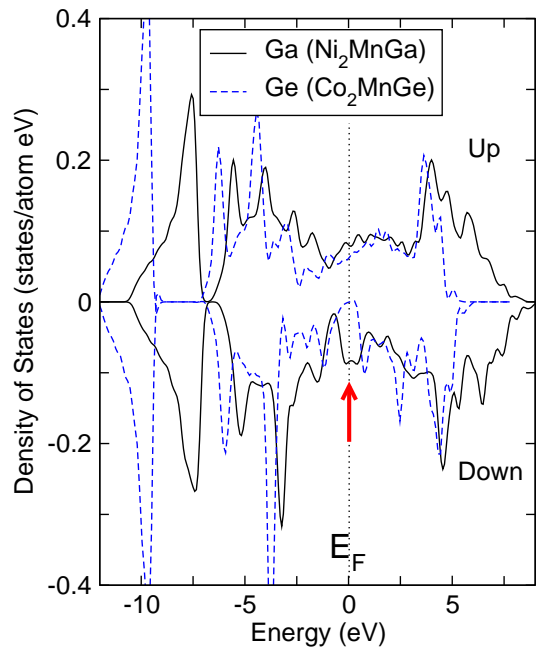}
  \end{center}
  \caption{(Color online) Site projected electronic density of states
  of Ni$_2$MnGa 
  and Co$_2$MnGe in the L2$_1$ structure. There is a difference in 
  the density of spin-down electronic states of both compounds close 
  to the Fermi level. In Ni$_2$MnX (X = Ga, Ge, Al, In) compounds, 
  the peak in the DOS marked by the arrow arises from $p$-states of 
  the X atom. The double structure of this peak is related to the 
  hybridization splitting cause by the interaction of 
  4$p_{\downarrow}$ and Ni-3$d_{\downarrow}$ states \cite{Ayuela-02}.} 
  \label{dos}
\end{figure}

In order to understand this observation we
have to recall that Heusler alloys are very unusual metallic
 systems due to the presence of the so called $p$-elements 
(Ga,Al,Ge etc), which form partially filled bands close to
E$_F$. These bands allow the $d$ electrons of transition metals 
hybridize with the $p$ electrons of the $p$ element \cite{SP-Galanakis}. 
In case of Ni$_2$MnGa, the Ga atoms form energetically favorable 
hybrid states with Ni. These states give a peak in the spin-down
electronic density of states right at the Fermi level (see
Fig.~\ref{dos}). The presence of this peak around E$_F$ is related
to the anomalous vibrational properties of the optical mode T$_{2g}$.

In contrast to this, Co$_2$MnGe does not 
show a peak in the DOS close to E$_F$. The same kind of hybridization
probably occurs in this case, but the $p$ band in this case sits at
lower energies so that when $d$ electrons of Co enter this band we
observe what is called half-metallic behavior, i.e. there are no
spin-down states at the Fermi level as can be seen in Fig.~\ref{dos}.

%%%%%%%%%%%%%%%%%%%%%%%%%%%%%%%%%%%%%%%%%%%%%%%%%%%%%%%%%%%%%%%%%%%%%
%%%%%%%%%%%%%%%%%%%%%%%%%%%%%%%%%%%%%%%%%%%%%%%%%%%%%%%%%%%%%%%%%%%%%%

\begin{figure} % Fig. 12
  \begin{center}  
\includegraphics*[angle=0,width=7.0cm]{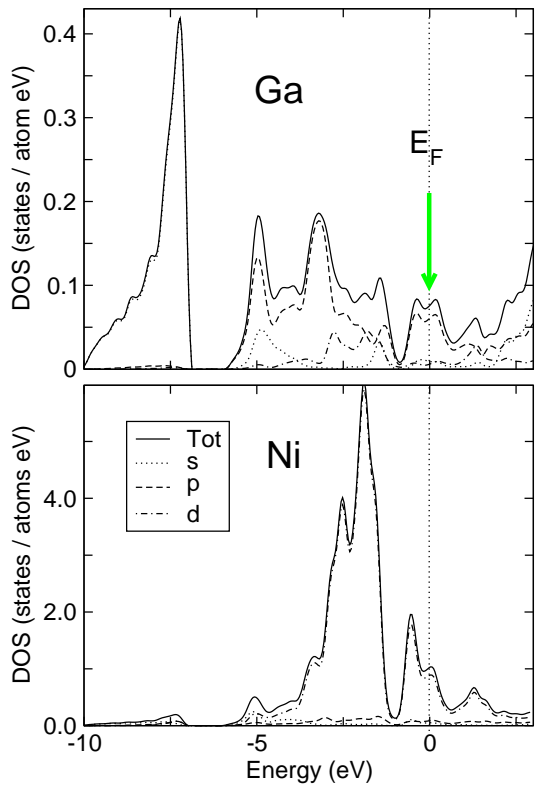}
  \end{center}
  \caption{(Color online) Site projected electronic density of states
  of Ni$_2$TiGa.  
  The double peak structure at the Fermi level is caused by the 
  hybridization of Ga 4$p$ and Ni 3$d$ states showing that  
  non-magnetic Heusler systems can also exhibit the same kind of 
  characteristic peak structure as the FM systems.} 
  \label{para_dos}
\end{figure}

This hybridization features affect binding mechanisms in Heusler alloys
to a large extent. These intermetallics and a competition of covalent
interaction and magnetic ordering was already discussed by K\"ubler 
\textit{et al.} in \cite{Kuebler83}. In our case, the covalent
bond stems from the 4$p$ electrons of Ga which can couple to the 
3$d$ electrons of Ni.
The magnetism is mainly governed by the
magnetic Jahn-Teller effect. This effect splits the giant peak of Mn 3$d$
states, which is present in the non-magnetic DOS at the Fermi 
level \cite{Veliko-99}. The stability of the FM Heusler systems
depends on a balance between magnetic ordering of the Mn atoms
and covalent bonding of Ga and Ni atoms.
However, we have shown, taking Ni$_2$TiGa as an example, that 
non-magnetic Heusler compounds exhibit the same kind of martensitic
instability as the FM ones and the feature in DOS is also present in
this case (see Fig.~\ref{para_dos}). Therefore, the magnetic order cannot be 
considered to be at the origin of the structural stability. Instead, 
more general features of the band occupations and the hybridization
have to be considered. It is important that while in metallic systems
the covalent interactions are expected to be weak, the present
calculations of the phonon dispersions show that some Heusler systems 
can exhibit strong anomalies due to the covalent bonding.

%%%%%%%%%%%%%%%%%%%%%%%%%%%%%%%%%%%%%%%%%%%%%%%%%%%%%%%%%%%%%%%%%%%%%%
%%%%%%%%%%%%%%%%%%%%%%%%%%%%%%%%%%%%%%%%%%%%%%%%%%%%%%%%%%%%%%%%%%%%%%

It is important that the low lying optical modes of Ni are the only
Raman-active modes. We know from the mechanism of
Raman scattering that this process must involve the polarizability of
the material. This is certainly related to the covalent interaction
between the Ga 4$p$ and Ni 3$d$ states. The distortion of the electron cloud
will help us further to consider one more aspect of this issue because 
the tetrahedral environment of the atoms in the Heusler alloys leads
to a natural distortion of the charge around X atoms (in X$_2$YZ; X=Ni,Co,Fe).

Apparently, it is difficult to obtain experimental evidence of
the optical modes inversion since the metallic character of our
systems will wash-out any signals of the Raman activity. However, 
Raman-active Ni vibrations have so far been observed in an amorphous metal
\cite{lustig-1985}. A Raman scattering experiment
on Ni$_2$MnGa would allow to follow up the vibrational behavior of Ni from
the parent phase through the martensitic structure with decreasing
temperature.
We suggest that the anomalous sequence of the optical phonons in 
the structurally unstable Heusler systems is a general feature and 
can be applied generally to other intermetallics.

 Finally, since we have involved the issue of optical phonons and
covalent interactions in our
discussions, it is worth thinking here about possible similarity with 
ferroelectrics, in which the structural instability can also be 
discussed on the basis of hybridization effects, in this case of 
oxygen $p$ and transition metal 3$d$ states. For the unusual behavior 
of phonon dispersion in ferroelectrics see, for example, 
\cite{ghosez-1999}.

%%%%%%%%%%%%%%%%%%%%%%%%%%%%%%%%%%%%
\subsection{Trends with e/a}
%%%%%%%%%%%%%%%%%%%%%%%%%%%%%%%%%%%%

The parameter $e/a$ is of rather qualitative nature and
has to be discussed in relation to other details of the electronic
structure discussed above.   
Here, we present some observations which can be useful in
the development of a conceptual macroscopic view on the stability of
Heusler structures.

The character 
of softening of the acoustic TA$_2$ mode of Ni$_2$MnGa, Ni$_2$MnAl and 
Ni$_2$MnIn is similar. For these compounds $e/a$ is the same,  $7.5$.
If we compare the phonon dispersions of these systems to those having one
additional valence electron, i.e.\ $e/a=7.75$, as in the case of
Ni$_2$MnGe, we find that the onset of instability at $\zeta \approx
0.6$ is more pronounced. 
We do not focus here on the additional instability in the
dispersion of Ni$_2$MnGe at $\Gamma$, it has a different origin.
In contrast to this unstable behavior of the Ni-based systems, the 
phonon dispersions of ferromagnetic Co-based stoichiometric compounds
do not show any trace of a structural instability 
(Fig.\ \ref{g1}). In this case, the number of valence electrons is 
smaller than $7.5$. Thus, we see that Co- and Ni- based systems show
different behavior with respect to the shear instability involving
the TA$_2$ phonon mode. Although, it is not clear what is that key
difference between Co and Ni, which makes Heusler structures stable
and unstable. Intrinsic properties of the Ni and Co atoms
would require different kind of approach not used in this paper.
It is left for future discussions in order to concentrate on other
aspects of our work.   
   
\begin{figure} % Fig. 9
  \begin{center}  
 \includegraphics*[angle=0,width=7.0cm]{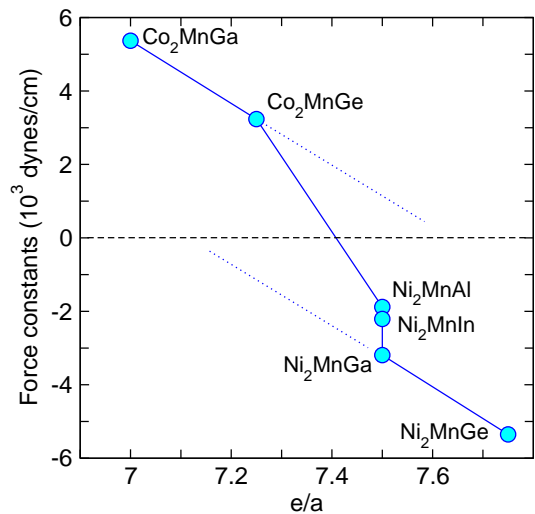}
  \end{center}
  \caption{(Color online)
  Nearest neighbor Ni-Ni and Co-Co force constants (X atoms in X$_2$YZ 
  Heusler compounds). The force constants strongly depend on the 
  valence electron number per atom. Possibly, there is a critical value of 
  $e/a$ allowing the separation of structurally unstable Heusler systems 
  with negative force constants from those of stable ones with positive 
  force constants, but this has to be verified by additional
  calculations. Note that these Ni-Ni negative force constants do 
  not lead directly to the unstable acoustic mode TA$_2$, but it
  leads to the inversion of the optical modes.} 
  \label{shift}
\end{figure}

We come back to the value of $e/a$ in order to check if there is a trend
which suggests a critical $e/a$ value 
to determine whether the L2$_1$ phase is stable or unstable. Maybe it
is possible regardless what kind of atom (Ni of Co) we have in the
structure, which would be very useful for analizing alloying effects
in the Heusler alloys and their stability. 
For this purpose we consider nearest neighbor force constants between
the atoms sitting in the 8c Wyckoff positions i.e. Ni or Co in our
case. Figure \ref{shift} presents results for only ferromagnetic
systems. From this picture we can see that both for Ni- and Co- based
systems the force constants follow similar trends suggesting lost of
their stability at some critical values for $e/a$. Here we have two
possibilities.  
We extrapolated lines corresponding Co- and Ni- based systems and show
that thay have different behavior. Such an approach shows that
replacing Co by Ni would shift the linear dependence down and other way
around if we repkace Ni by Co. Contrary to 
that, we can assume that the kind of atom is not important, but the
$e/a$ dependence of the force constants is not linear and a
crossover from stable to unstable configurations occurs at around $e/a
= 7.4$. However, our data is not sufficient to draw any final
conclusion for these two scenarios. This needs to be verified in
future studies.   

We would like to note that when looking at three systems,
Ni$_2$MnAl, Ni$_2$MnGa and Ni$_2$MnIn with the same $e/a = 7.5$ we
find that the differences in their force constants are
small. This shows that the size and mass of Al, Ga or In do not play
much role, but the similar valence electron filling seems to be more
important.

%%%%%%%%%%%%%%%%%%%%%%%%%%%%%%%%%%%%%%%%%%%%%%%%%%%%%%%%%%%%%%%%%%%%
%%%%%%%%%%%%%%%%%%%%%%%%%%%%%%%%%%%%%%%%%%%%%%%%%%%%%%%%%%%%%%%%%%%%

We did not include Ni$_2$TiGa and Fe$_2$MnGa in the Fig.\ref{shift}
keeping in mind that the type of magnetic order should lead to 
significant distinctions in how the electrons fill
appropriate energy levels with possible relation to the critical ratio
of $e/a$.
With respect to the role of the magnetic order,
it is helpful to keep in mind previous results
related to the interplay of magnetic order and martensitic instability has
previously been discussed for Ni$_2$MnGa 
\cite{Planes-97,Harmon}. It has been shown that 
the spin-split band structure gives rise to a prominent peak in the
the wave vector dependent magnetic susceptibility close to 
${\bf q} = \frac{2\pi}{a} \, (\frac{1}{3},\frac{1}{3},0)$,
which corresponds to the experimentally observed wave vector for the
premartensitic instability \cite{Harmon}. This enhanced susceptibility
has its origin in Fermi surface nesting in the different spin
channels. The fact that FM Co-based systems are stable,
although tendencies of Fermi surface nesting exist, reflects an
obvious situation when
the characteristic topology of the Fermi surface must
strongly depend on the  $e/a$ ratio.

%%%%%%%%%%%%%%%%%%%%%%%%%%%%%%%%%%
\subsection{Tetrahedral coordination of the covalent bonding}
%%%%%%%%%%%%%%%%%%%%%%%%%%%%%%%%%%

\begin{figure} % Fig. 13
  \begin{center}  
\includegraphics*[angle=0,width=6.0cm]{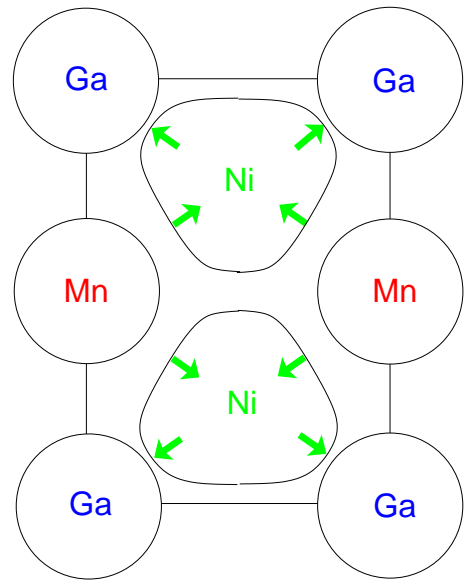}
  \end{center}
  \caption{(Color online)
  Schematic representation of the deformation of the outer 
  3$d$ electron cloud around the Ni atoms in the unstable Heusler 
  alloys from present \textit{ab initio} calculations. 
  There are two effects which contribute to the deformation. One is due to
  covalent bonding of Ga 4$p$ and Ni 3$d$ states, the other (stronger) 
  effect is due electrostatic repulsion of Mn and Ni atoms
  resulting from their nearly spherical charge distribution. 
  The deformation of the 3$d$ orbitals of Ni is of tetrahedral
  nature and strengthens the covalent Ni-Ga bond.}  
  \label{push}
\end{figure}

In order to further elucidate the role of the covalent interaction, we have 
analyzed the distribution of the electronic charges in our
ferromagnetic Heusler alloys. 
The comparison of the spin-up and spin-down densities of states  
has shown a clear difference in the charge distribution. 
The issue of the electronic hybridization becomes clear if
we analyze the situation in Ni$_2$MnGa by using a simple model. Figure
\ref{push} shows schematically the charge distribution (spin-down
only) taken from \textit{ab initio} calculations for Ni$_2$MnGa in the
(110) plane of the tetragonal unit cell shown in Fig.\ \ref{struc}b. 
We observe an electrostatic-like repulsion between the Mn and Ni
atoms. The reason for this effect is the large magnetic moments of
Mn. Spin-down 3$d$ electrons of Ni do not find symmetry-allowed
$d$-states of Mn to hybridize with, unless they flip their
spin. Instead, the 3$d$ electrons of Ni  hybridize with the
4$p$ electrons of Ga. It looks like the charge of Ni is pushed 
towards the neighboring Ga atoms, which are tetrahedrally 
coordinated around it. Thus, the tetrahedral coordination which we
find in the Heusler alloys is a 
natural basis for the 3$d$ electrons of Ni to hybridize with the
4$p$ electrons of Ga.

%%%%%%%%%%%%%%%%%%%%%%%%%%%%%%%%%%%%%%%%%%%%%%%%%%%%%%%%%%%%%%%%%%%%%%
%%%%%%%%%%%%%%%%%%%%%%%%%%%%%%%%%%%%%%%%%%%%%%%%%%%%%%%%%%%%%%%%%%%%%%

%%%%%%%%%%%%%%%%%%%%%
\section{Conclusions}
%%%%%%%%%%%%%%%%%%%%%

Our total energy density functional theory investigations of the 
structural instabilities in non-magnetic and magnetic Heusler
compounds allow us to draw following conclusions. 

We have found that in all Heusler compounds which show acoustic
unstable mode TA$_2$, the optical vibrations exhibit unusual
inversion of their modes. Analysis of the force constants revealed that
negative force constants are present in all unstable systems, but these
force constants are not directly related to the acoustic phonon
instability. Instead, the inversion of the optical modes appears to be
the driving force for the acoustic anomaly. The Raman-active optical
modes lower their energy due to the negative force constants, whereby
the acoustic mode TA$_2$ becomes unstable due the repulsive
interaction with the anomalous optical mode. This acoustic-optical
interaction might be of more general interest then just for the
Heusler compounds. Its role in the martensitic transformations has to
be investigated in more detail, including those in binary alloys.

A comparative analysis of different systems  
has shown that unstable Heusler compounds show 
characteristic features in the electronic DOS. The peak in the 
electronic states appears at the Fermi level, being  
responsible for the nesting topology of the Fermi surface of the
unstable Heusler systems and the covalent bonding features which
affect the optical vibrational modes of the Ni-based systems. 

We obtained that phonon instabilities observed in the Heusler
compounds can be related to the number of valence electrons in the
systems. However, our consideration of only Co- and Ni-based
systems does not allow us to discuss how general this observation can
be. A crossover from the stable Co- to unstable Ni-based systems on
the basis of the $e/a$ value can be suggested, but additional
investigations are required. Microscopic intrinsic properties of Co
and Ni atoms may stand for a substantial part of our observations.

Study of additional Heusler compounds (see Ref.~\onlinecite{LB32c})
will serve to further establish physical trends to the problem of
martensitic phase transitions.

%%%%%%%%%%%%%%%%%%%%%%%%%%
\section{Acknowledgements}
%%%%%%%%%%%%%%%%%%%%%%%%%%

This work was supported by SFB-491 (Germany) and NSF MRSEC
DMR-00-80008 (U.S.A.). ATZ thanks E. Liskova, K. Parlinski,
A. Postnikov and C. Fennie for very helpful and inspiring discussions. 

%%%%%%%%%%%%%%%%%%%%%%%%%%%%%%%%%%%%
{\small
  \bibliographystyle{apsrev}
  \bibliography{zayak-entel-heusler}
}
%%%%%%%%%%%%%%%%%%%%%%%%%%%%%%%%%%%%

%%%%%%%%%%%%%%
\end{document}